\documentclass[journal]{IEEEtranTIE}
\usepackage{graphicx}
\usepackage{cite}
\usepackage{picinpar}
\usepackage{amsmath}
\usepackage{url}
\usepackage{flushend}
\usepackage[utf8]{inputenc}
\usepackage{colortbl}
\usepackage{soul}
\usepackage{multirow}
\usepackage{booktabs}
\usepackage{pifont}
\usepackage{color}
\usepackage{alltt}
\usepackage[hidelinks]{hyperref}
\usepackage{enumerate}
\usepackage{siunitx}
\usepackage{breakurl}
\usepackage{epstopdf}
\usepackage{pbox}
\usepackage{tikz}
\usepackage{circuitikz}
\usetikzlibrary{shapes.geometric, arrows}
\usepackage{xcolor}
\definecolor{darkgreen}{rgb}{0,0.5,0} % custom dark green
\usepackage{booktabs}
\usepackage{tabularx}
\usepackage{subcaption}
\usepackage{pgfplots}
\pgfplotsset{compat=1.18}

\usepackage{comment}

\tikzstyle{startstop} = [rectangle, rounded corners, text centered, draw=black]
\tikzstyle{process} = [rectangle, text centered, draw=black]
\tikzstyle{arrow} = [thick,->,>=stealth]

\begin{document}

\title{
    % Modular STATCOM with Fully Sensorless Module Balancing and Minimum Transistor Count
    % Modular STATCOM with Fully Sensorless Module Balancing Based on Four-Transistor Four-Diode (4T4D) Series/Parallel Chopper Cell
    % Modular STATCOM with Sensorless Module Balancing Based on Four-Transistor Four-Diode (4T4D) Series\,/\,Parallel Chopper Structure
    Four-Transistor Four-Diode (4T4D) Series/Parallel Chopper Module for Auto-Balancing STATCOM and Low Control and Development Complexity 
    % Direction-Selective Parallel Modular Static Compensator with Sensorless Voltage Balancing and Minimum Transistor Count
    % Sensorless Modular STATCOM with Hardware-Guaranteed Voltage Balancing Using the Direction-Selective Parallel Topology
    % Direction-Selective Parallel Topology for Modular STATCOMs: Minimum Switch Count with Sensorless Voltage Balancing
    % Modular STATCOM with Bidirectional Parallelization and Sensorless Voltage Regulation Using Four Transistors per Module
    % Sensorless Voltage-Balanced Modular STATCOM Based on Direction-Selective Parallel Modules with Minimum Transistor Count
}

\author{
	\vskip 1em
	Jinshui Zhang, \emph{Student Member},
	Zane Mannings, 
	Chris Dittmer, 
	Angel V Peterchev, \emph{Fellow},
	and Stefan M Goetz, \emph{Member}
    % \thanks{All authors are with Duke University, Durham, NC, USA.}
}

\maketitle
\begin{abstract}
    Static synchronous compensators (STATCOMs) manage reactive power compensation in modern power grids and have become essential for the integration of renewable energy sources such as wind farms. 
    Cascaded H bridges have become the preferred topology for high-power STATCOMs, but balancing module capacitor voltages remains a persistent challenge. 
    Conventional solutions equip every module with a voltage sensor---a component that is costly, temperature-sensitive, and prone to aging-related failures.
    Recent parallel-capable module topologies can balance voltage through switched-capacitor operation. 
    The latest developments reduced the sensor requirement from one per module to one per arm. 
    However, these implementations require twice as many individual transistors compared to series-only topologies.
    We present a STATCOM solution based on the four-transistor four-diode (4T4D) series\,/\,parallel chopper cell. This topology achieves bidirectional parallelization with only four transistors per module---exactly as many as a conventional full bridge.
    Furthermore, we propose a dual-loop control strategy that fully eliminates module voltage sensors by inferring voltage levels from the modulation index.
    This scheme also improves output quality by regulating the modulation depth.
    We validated our proposal through simulation and experiments.
    We built a prototype to interface the grid. The prototype further passed robustness tests with step change, current direction reversal, and grid disturbance. 
    This work demonstrates the first modular STATCOM implementation that combines minimum transistor count with complete elimination of module voltage sensors.
\end{abstract}


% \markboth{IEEE TRANSACTIONS ON INDUSTRIAL ELECTRONICS}%
{}
\section{Introduction}

The concept of the flexible AC transmission system improves the controllability, stability, and power transfer capacity of modern grids \cite{hingorani2002flexible, peng2017flexible}. Among these, the static synchronous compensator (STATCOM) can dynamically source or sink reactive power and distortion to support the grid voltage and correct the power factor \cite{singh2009static, sharma2023comprehensive}. Compared with traditional passive compensation devices that switch capacitor or inductor banks, STATCOMs offer faster response, continuous control, and bidirectional reactive power flow. 

% The development of high-power semiconductor devices such as gate turn-off thyristors and insulated-gate bipolar transistors (IGBTs) in the 1980s enabled practical voltage-source converter implementations for STATCOM applications. Early STATCOMs employed two-level converter topologies, which produce square-wave outputs, and thus require substantial filtering inductors and exhibit poor performance at light loads \cite{singh2009static}. Three-level topologies, such as the neutral-point-clamped converter, reduce harmonic distortion by adding an intermediate voltage level \cite{saeedifard2007space, sharma2023comprehensive}. Multi-pulse configurations use phase-shifting transformers to further improve output quality but at the cost of bulky magnetics \cite{maurya2020detailed}. In high-voltage applications, these monolithic topologies require series connection of multiple switching devices. The consequential challenges of gate drive design and transistor voltage sharing collectively motivated the adoption of modular cascaded converter topologies \cite{bina2010transformerless}.

Cascaded bridge circuits (CBCs)\,/\,modular multilevel converters (MMCs) have become the preferred solution for STATCOMs, as they distribute the voltage stress and power load across multiple identical modules \cite{tang2016compact}. Each module contains a dc capacitor and a set of semiconductor switches that insert or bypass the capacitor voltage. The number of output voltage levels scales with the number of modules, which enables high-quality waveforms without excessive filtering. The modular architecture also provides inherent redundancy---faulty modules can be bypassed while the system continues operation.

One of the most critical issues in MMCs\,/\,CBCs is module voltages drifting apart when individual charge flows are not perfectly balanced over time. This voltage deviation can be influenced by current direction, amplitude, and capacitor as well as semiconductor leakage tolerances. Unequal module voltages entail uneven output steps and waveform distortion. Severe imbalance may cause excessive voltage and permanently damage the hardware \cite{merlin2015cell}.

An extensive body of research has focused solely on voltage balancing through control algorithms \cite{wu2021comprehensive, eroglu2022critical, milovanovic2021comprehensive, konstantinou2018estimation, dekka2017evolution, ma2018enhanced}. However, most schemes require every module to be equipped with a voltage sensor---a component that is temperature-sensitive, expensive, and often fails to guarantee hardware-safe operation independent of software. Software, however, is considered a fault-prone element by functional safety standards.

One solution circumventing these sensor-related risks is to introduce additional parallel connectivity between modules. Instead of bypassing inactive modules, parallel-capable topologies can temporarily connect them to neighboring modules, which establishes direct charge equilibrium \cite{goetz2015modular, fang2021review, goetz2016sensorless, fang2021multilevel, li2019module}. While paralleled, the module group can act as a single unit with larger effective capacitance, lower source impedance, and reduced voltage ripple.

The alternation between series and parallel connectivity between modules and therefore module capacitors is a powerful operating mode. It enables switched-capacitor behavior in cascaded-bridge converters. The switched-capacitor behavior is concurrent with the conventional cascaded-bridge operation.

Most importantly, however, the parallel connectivity drastically simplifies balancing. It achieves ideal voltage maintenance and charge sharing as a hardware solution \cite{li2019module, goetz2017control, zhang2024frequency}. Balancing can run open loop without error-prone software control or any module voltage sensors.

A growing number of topologies have been proposed to enable parallel connectivity in MMCs\,/\,CBCs \cite{fang2021review}. 
The cascaded double-H-bridge (CH$^2$B) topology offers the most complete functionality with bidirectional parallelization and bipolar output capability, but at the cost of eight transistors per module \cite{goetz2015modular}. Symmetrical and asymmetrical double-half-bridge variants halve the switch count but sacrifice output polarity or double the number of capacitors \cite{fang2021reduction, li2019module}. Diode-clamped and switch-clamped modules provide simplified structures with unidirectional parallelization paths \cite{tashakor2020modular, yin2018modular, liu2016novel, zheng2017medium, tashakor2021switch, tashakor2023voltage}; however, these topologies cannot guarantee ideal voltage balancing without active control assistance.

Recent modular STATCOM systems exploit those features of a parallel-enabled cascaded topology \cite{li2021reduced}. This solution reduces the sensor count from one per module to one per arm. Although effective, this approach requires a larger number of individual transistors compared to half or full bridges. The transistor current utilization is as high as in conventional bridges so that the total amount of silicon is the same, but the many individual transistors also need individual gate driving and control.

This paper presents a STATCOM solution with series and parallel inter-module connectivity but only exactly as many transistors as a conventional H bridge. First members of this class of converters were recently presented \cite{zhang2024direction}. These circuits achieve bidirectional parallelization and bipolar output with only four transistors per module, which is equal to a standard full-bridge without parallel connectivity.

In addition, we develop a dual-loop control strategy that completely eliminates module voltage sensors. The inner loop regulates the reactive output current, while the outer loop controls the modulation index amplitude and consequently the module voltages.
These innovations together yield an ideally balanced modular STATCOM with four transistors, simple balancing, and zero voltage sensors per module.

\section{Hardware Design}
\begin{figure*}[t]
    \centering
    \includegraphics{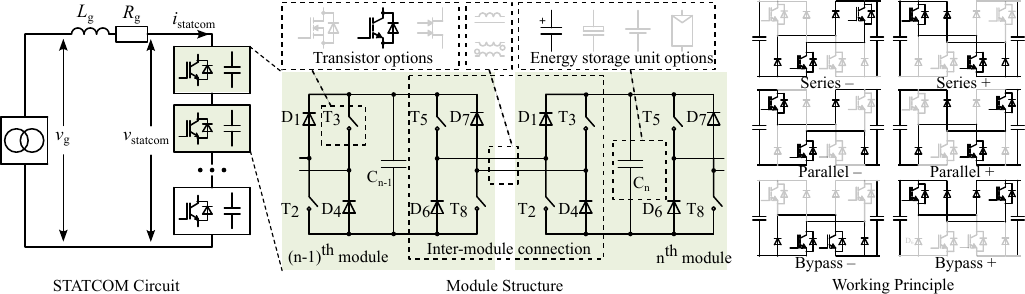}
    \caption{A modular STATCOM circuit featuring the four-transistor four-diode chopper (4T4D) series\,/\,parallel topology.}
    \label{fig:statcom_disep_illustrator}
\end{figure*}

\begin{table*}[t]
\centering
\caption{Comparison of MMC Module Topologies for STATCOM Applications}
\label{tab:topology_comparison}
\begin{tabular}{l c c c c c}
\toprule
Module Topology & Transistor Count & Capacitor Count & Output Polarity & Parallelization & Fault Blocking \\
\midrule
Half-bridge modules (purely series) & 2 & 1 & Unipolar & None & No \\
Full-bridge modules (purely series) & 4 & 1 & Bipolar & None & Yes \\
Double half bridge modules (series\,/\,parallel) \cite{li2019module} & 4 & 1 & Unipolar & Bidirectional & No \\
Diode-clamped modules (passive parallelization backbone) \cite{tashakor2020modular} & 2 & 1 & Unipolar & Unidirectional & No \\
Switch-clamped modules (active parallelization backbone) \cite{tashakor2021switch} & 3 & 1 & Unipolar & Unidirectional & No \\
Symmetric double half bridge (series\,/\,parallel) \cite{fang2021reduction} & 4 & 2 & Bipolar & Bidirectional & No \\
Cascaded double H bridge (series\,/\,parallel) \cite{goetz2015modular} & 8 & 1 & Bipolar & Bidirectional & Yes \\
\textbf{4T4D series\,/\,parallel chopper cell (proposed)} & \textbf{4} & \textbf{1} & \textbf{Bipolar} & \textbf{Bidirectional} & \textbf{Yes} \\
\bottomrule
\end{tabular}
\end{table*}

\subsection{Topology}
Table~\ref{tab:topology_comparison} compares the key characteristics of various module topologies for cascaded structures. The half-bridge and full-bridge (H bridge) modules represent conventional designs without parallel connection; they rely entirely on active software-based balancing control. The CH$^2$B topology introduced hardware-guaranteed balancing through bidirectional parallelization but requires eight individual transistors and gate drivers per module, though the same amount of power silicon as H bridges. Symmetrical and asymmetrical double-half-bridge variants reduce the number of individual transistors but lose negative output polarity or double the capacitor count.

Diode- and switch-clamped topologies simplify structures with fewer transistors and a hardware balancing path through parallelization. However, this parallelization is purely unidirectional and cannot easily implement other features known from cascaded series\,/\,parallel converters. The new 4T4D chopper topology offers all states of other cascaded series\,/\,parallel modules, such as the double H bridge, but needs only four transistors. In contrast to diode-clamped cascaded bridges, as another option with a small transistor count, it allows full switched-capacitor energy exchange in both directions.  

Figure~\ref{fig:statcom_disep_illustrator} illustrates a STATCOM with 4T4D cells. Each module comprises one capacitor and two types of chopper half-bridges: Two diode--switch ($D_1$--$T_2$ and $D_7$--$T_8$) and switch--diode ($T_3$--$D_4$ and $T_5$--$D_6$) choppers require four transistors and four diodes. We use bipolar power transistors, specifically insulated-gate bipolar transistors (IGBTs), which match the diodes well in dynamics and loss behavior. Two adjacent modules are connected through two leads, which transport both the load current (common mode) and the (switched-capacitor) balancing current (differential mode). As in other series\,/\,parallel cascaded module structures, the module interconnection can include compact magnetics, also selectively for common- or differential-mode currents to tune parallelization dynamics or distribute output filtering \cite{10721336}.

\subsection{Working Principles}

The 4T4D module supports six active module-interconnection states (Figure \ref{fig:statcom_disep_illustrator}). The two \textit{Series} states provide negative and positive series connection of the capacitors of adjacent modules. The current flows in parallel through two paths, each with a diode and a transistor for low loss and high semiconductor utilization. The two \textit{Parallel} states temporarily parallelize the capacitors of the involved modules for energy exchange and can enforce the power-flow direction; the first parallel state only allows power flowing in one direction, the second in the opposite direction. The two \textit{Bypass} states conduct the load current along either the positive or the negative rail. A bypassed module does not increase nor reduce the output voltage.
In the \textit{Passive} state, the transistors are off, and the module acts as a rectifier. Like other cascaded bridges, it can still control and contribute to the output voltage but absorbs output power.

\subsubsection{Direction-Selective Parallelization}
The circuit contrasts with other cascaded series\,/\,parallel converters, such as double H-bridge or double half-bridge circuits, in an important detail. In known series\,/\,parallel modules, the parallel module state equilibrates paralleled capacitors in either direction. 4T4D cells have two parallel states, which selectively only allow differential-mode current and power flow in one of the two directions. In the \textit{Parallel$-$} mode, capacitor $C_{n-1}$ charges $C_n$, whereas the opposite occurs in the \textit{Parallel$+$} mode. These two modes offer complementary parallelization directions, which underlies the name \textit{direction-selective parallel}.

A control method can exploit the two selective power-flow directions deliberately. If this feature is not needed, the controller can simply alternate between \textit{Parallel$+$} and \textit{Parallel$-$} for bidirectional energy exchange. A simple rising-edge latch and output status register can alternate the two parallel modes in the modulation scheme.

\subsubsection{Self-Equalizing Behavior and Reduced Paralleling Loss}
The bipolar devices in 4T4D cells generate a natural equalization corridor, which suppresses any oscillations, limits the voltage difference between modules to a small band, and reduces loss.
Minor voltage mismatches that fall below this threshold do not trigger a parallelization event and therefore produce no paralleling loss. 

\subsubsection{Impedance Matching and Transistor Design}
In the \textit{Bypass} and \textit{Parallel} states, 4T4D modules have the same equivalent impedance as conventional H-bridges---one transistor and one diode in series. In the \textit{Series} modes, current flows through two parallel branches, each with one transistor and one diode. 
IGBTs provide good impedance matching with the diodes. Furthermore, IGBTs typically offer better over- and fault-current capability compared to unipolar field-effect transistors.

Field-effect transistors may typically outcompete IGBTs in loss at lower voltages, fast switching rates, and low currents. Cascaded bridges in medium- and high-voltage applications, such as most STATCOMs, typically have to manage grid fault currents, use high module voltages, where the pn forward voltage becomes less relevant, and operate at low to moderate module switching rates; they still achieve a high effective system-level switching rate with low loss \cite{singh2009static,AMIRREZAI2023109253}.

\section{Control}
\begin{figure*}
    \centering
    \includegraphics{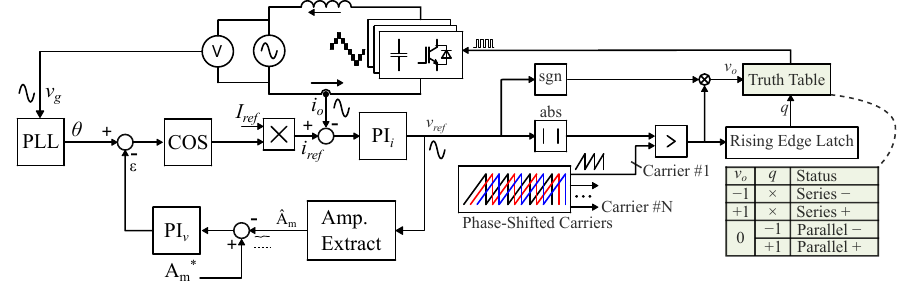}
    \caption{Dual-loop control structure for the 4T4D topology based STATCOM. The inner current loop controls reactive current injection; its PI output drives phase-shifted-carrier PWM to synthesize the multilevel output voltage. An additional rising-edge latch alternates the two parallel modes. The outer loop regulates the modulation depth toward a target value $A_\textrm{m}^*$, which implicitly sets the module capacitor voltage; its PI output produces a phase offset $\varepsilon$ that adjusts the active power exchange. }
    \label{fig:control_diagram}
\end{figure*}

\subsection{Voltage Balancing Strategies in Series\,/\,Parallel MMC}

Conventional modular STATCOMs employ a dual-loop control architecture. The inner current loop tracks a sinusoidal reactive current reference by adjusting the converter output voltage. The outer voltage loop measures individual module capacitor voltages, compares them against a common set point, and adjusts the active power exchange with the grid to compensate converter losses and maintain voltage balance.

Series\,/\,parallel modular multilevel circuits, such as those based on CH$^2$B and also the new 4T4D structure, achieve local voltage equalization through hardware parallelization.
Previous STATCOMs with such topologies still require one voltage sensor per arm to form a regulation loop \cite{li2021reduced}. This sensor measures a representative module voltage to detect overall drift in the arm's energy storage. Despite the reduction of sensors, this approach now depends critically on a single measurement point per arm. A single sensor failure, whether due to temperature drift, aging, or malfunction, can compromise the entire arm's operation rather than affecting only one module. The inherent reliability concerns of voltage sensors persist.

\subsection{Elimination of Module Voltage Sensors through Modulation Depth Regulation}
We designed a two-stage control that differs from the prior art. 
The inner loop controls the current. As the module design allows hardware-level open-loop balancing of the module capacitor voltages, we could perform additional active balancing, but it is not necessary.
We therefore actively control the average module voltage in the outer loop to achieve a certain modulation index $A_\textrm{m}$, i.e., to maximize the output voltage granularity with a suitable module step size.

The feedback signal $\hat{A}_\textrm{m}$ is extracted from the amplitude of the inner current PI loop's output. A dedicated additional sensor is not necessary. The outer loop regulates $\hat{A}_\textrm{m}$ toward a designer-chosen target $A_\textrm{m}^*$ by adjusting the active power exchange with the grid. The module capacitor voltage is not a setpoint; it is an emergent consequence that settles at a value that can satisfy the target modulation depth.

This design confers two distinct advantages over the conventional approach. First, sensor elimination follows naturally: $\hat{A}_\textrm{m}$ is derived from an already-computed signal. Second, the choice of $A_\textrm{m}^*$ directly governs the converter's steady-state modulation depth, which determines output waveform quality. A higher $A_\textrm{m}^*$ drives the converter to employ more active output voltage levels and thereby reduces harmonic distortion. The control of the modulation depth gives the designer an explicit handle on output waveform quality. This degree of freedom is absent in conventional fixed-voltage-setpoint schemes, where the modulation depth varies passively with grid conditions.

The steady-state voltage at the STATCOM terminal is modeled by Kirchhoff's voltage law per
\begin{equation}
    \mathbf{v}_\textrm{o} = \mathbf{v}_\textrm{g} + j\omega L_\textrm{g} \mathbf{i}_\textrm{o} + R_\textrm{g}\mathbf{i}_\textrm{o},
    \label{eq:vstatcom_kvl}
\end{equation}
where $\mathbf{v}_\textrm{g}$, $\mathbf{v}_\textrm{o}$, and $\mathbf{i}_\textrm{o}$ are complex phasors respectively representing the grid voltage, converter output voltage, and output current; $\omega$ is the angular grid frequency; and $L_\textrm{g}$ and $R_\textrm{g}$ are the grid-side inductance and resistance.

The converter synthesizes this voltage from $N$ series-connected submodules per
\begin{equation}
    \left| \mathbf{v}_\textrm{o} \right | = A_\textrm{m} \sum_{n=1}^N v_{\textrm{sm}, n} = A_\textrm{m} N V_{\textrm{sm}},
    \label{eq:vstatcom_modulation}
\end{equation}
where $A_\textrm{m}$ is the modulation index amplitude, $V_{\textrm{sm}}$ is the hardware-equalized module capacitor voltage.
% \begin{equation}
%     \left| \mathbf{v}_\textrm{o} \right | = A_\textrm{m} N V_{\textrm{sm}},
%     \label{eq:vstatcom_modulation}
% \end{equation}
% where $V_{\textrm{sm}}$ is the equalized module capacitor voltage.

The combination of \eqref{eq:vstatcom_kvl} and \eqref{eq:vstatcom_modulation} can regulate the modulation index to $A_\textrm{m}^*$, and the module voltage settles to
\begin{equation}
    \begin{aligned}
        V_{\textrm{sm}}^* &= \frac{\left| \mathbf{v}_\textrm{g} + j\omega L_\textrm{g} \mathbf{i}_\textrm{o} + R_\textrm{g}\mathbf{i}_\textrm{o} \right|}{A_\textrm{m}^* N} \\
        & \simeq \frac{\sqrt{\left(\mathbf{v}_\textrm{g} + 2\pi f_\textrm{g}L_\textrm{g} \mathbf{i}_\textrm{o}\right)^2 + \left(R_\textrm{g}\mathbf{i}_\textrm{o}\right)^2}}{A_\textrm{m}^* N}.
    \end{aligned}
    \label{eq:vsm_estimation}
\end{equation}
This equation shows that setting $A_\textrm{m}^*$ implicitly determines the average steady-state module voltage $V_{\textrm{sm}}^*$ through the prevailing grid voltage and output current.

\subsection{Modulation-Index Estimation}
Extraction of the current PI controller output amplitude is a core function of the proposed control scheme. We discuss how to implement it in both single-phase and three-phase settings.

\subsubsection{Single-Phase STATCOM}
We can approximate the current loop PI output as 
\begin{equation}
    v_\textrm{m}(t) = \textrm{PI}_\textrm{i}(i_\textrm{ref} - i_\textrm{o}) \approx \hat{A}_\textrm{m}\sin(\omega t + \theta_\textrm{o}).
\end{equation}
We use synchronous (I/Q) detection, which exploits the orthogonality of sinusoidal references, for real-time estimation of the modulation index amplitude. The following equations describe a digital implementation.

The current PI output can be approximated as 
\begin{equation}
    v_\textrm{m}[k] \approx \hat{A}_\textrm{m}\sin(\omega k T_s + \theta_\textrm{o}[k]), 
\end{equation}
where $T_s$ is the sampling step.

The PI output is multiplied by trigonometric signals synchronized to the grid to provide
\begin{equation}
\begin{aligned}
I[k] &= v_\textrm{m}[k] \cos(\omega k T_s + \theta_\textrm{g}[k]), \\
Q[k] &= v_\textrm{m}[k] \sin(\omega k T_s + \theta_\textrm{g}[k]), 
\end{aligned}
\end{equation}
where $\theta_\textrm{g}$ is the grid phase angle from the phase-locked loop. Expansion of these products yields dc components plus double-frequency terms as
\begin{equation}
\begin{aligned}
    I[k] &= \frac{\hat{A}_\textrm{m}}{2}\left[\sin(\theta_\textrm{g} - \theta_\textrm{o}) + \sin(2\omega kT_s + \theta_\textrm{g} + \theta_\textrm{o})\right]\!, \\
    Q[k] &= \frac{\hat{A}_\textrm{m}}{2}\left[\cos(\theta_\textrm{g} - \theta_\textrm{o}) - \cos(2\omega kT_s + \theta_\textrm{g} + \theta_\textrm{o})\right]\!.
\end{aligned}
\end{equation}
A first-order recursive low-pass filter extracts the dc components as
\begin{equation}
\begin{aligned}
\bar{I}[k] &= (1-\alpha)\bar{I}[k-1] + \alpha I[k], \\
\bar{Q}[k] &= (1-\alpha)\bar{Q}[k-1] + \alpha Q[k],
\end{aligned}
    \label{equ:lpf_iq}
\end{equation}
where $\alpha = 2\pi f_\textrm{c} / f_\textrm{s}$ sets the filter bandwidth $f_\textrm{c}$. The amplitude is then computed as
\begin{equation}
    \hat{A}_\textrm{m}[k] \approx 2\sqrt{\bar{I}[k]^2 + \bar{Q}[k]^2}.
\end{equation}
The filter bandwidth trades off noise rejection against response time. A bandwidth range of $f_{\textrm{grid}}/20$ to $f_{\textrm{grid}}/5$ can provide adequate smoothing while maintaining an acceptable dynamic response for modulation-index regulation.

\subsubsection{Three-Phase STATCOM}
The amplitude estimation considerably simplifies in three-phase systems. The Clarke transformation naturally decomposes balanced three-phase signals into orthogonal $\alpha$--$\beta$ components per
\begin{equation}
\begin{bmatrix} v_\alpha \\ v_\beta \end{bmatrix} = \frac{2}{3}
\begin{bmatrix} 1 & -\frac{1}{2} & -\frac{1}{2} \\ 0 & \frac{\sqrt{3}}{2} & -\frac{\sqrt{3}}{2} \end{bmatrix}
\begin{bmatrix} v_\textrm{a} \\ v_\textrm{b} \\ v_\textrm{c} \end{bmatrix}\!\!,
\end{equation}
where $v_\textrm{a}$, $v_\textrm{b}$, and $v_\textrm{c}$ are the three-phase PI controller outputs. The $\alpha$--$\beta$ components are inherently orthogonal and equivalent to the I and Q channels derived through synchronous detection. The amplitude is then directly obtained as
\begin{equation}
\hat{A}_\textrm{m} = \sqrt{v_\alpha^2 + v_\beta^2}
\end{equation}
without explicit quadrature multiplication or low-pass filtering. 

\subsection{Modulation-Index Regulation via Active Power Control}
The modulation depth is regulated through controlled active power exchange with the grid. In a STATCOM operating ideally at unity power factor (pure reactive current), the average active power flow is near zero, and module voltages remain constant, neglecting losses. In practice, converter losses cause a gradual decline in module voltage, which raises the required modulation depth and must be compensated by absorbing a small amount of active power from the grid.

The active power flow is controlled by adjusting the phase angle $\varepsilon$ between the converter output current and the grid voltage. The outer loop (PI$_\textrm{v}$ in Fig.~\ref{fig:control_diagram}) generates this phase offset based on the error between the estimated modulation amplitude $\hat{A}_\textrm{m}$ and the target $A_\textrm{m}^*$. For the same output voltage, a low $\hat{A}_\textrm{m}$ corresponds to a high module voltage, and vice versa. When $\hat{A}_\textrm{m} > A_\textrm{m}^*$, which indicates that module voltages have dropped below their operating point, the controller introduces a small phase lag ($\varepsilon < 0$) that shifts the current waveform to include an in-phase (active) component. This active component delivers power to the converter, which charges the module capacitors and reduces $\hat{A}_\textrm{m}$ back toward $A_\textrm{m}^*$. Conversely, when $\hat{A}_\textrm{m} < A_\textrm{m}^*$, a phase lead ($\varepsilon > 0$) causes the converter to export active power, which discharges the capacitors and increases $\hat{A}_\textrm{m}$.

The instantaneous active power of the STATCOM follows
\begin{equation}
    P = V_\textrm{g} I_\textrm{o} \cos(\theta_\textrm{g} - \theta_\textrm{i}) \approx V_\textrm{g} I_\textrm{o} \sin(\varepsilon),
\end{equation}
where $V_\textrm{g}$ is the grid voltage amplitude, $I_\textrm{o}$ is the output current amplitude, $\theta_\textrm{g}$ is the grid voltage phase, and $\theta_\textrm{i} = \theta_\textrm{g} + \pi/2 - \varepsilon$ is the current phase for a nominally reactive (leading or lagging by $90^\circ$) current with phase offset $\varepsilon$. For small $\varepsilon$, the active power is approximately linear in the phase offset, enabling straightforward PI controller design.

The outer modulation-index loop generates this phase offset based on the modulation index error as
\begin{equation}
\varepsilon = \textrm{PI}_\textrm{v}(A_\textrm{m}^* - \hat{A}_\textrm{m}).
\end{equation}
When module voltages drop, $\hat{A}_\textrm{m}$ increases to produce a negative $\varepsilon$ that introduces an in-phase current component to draw active power from the grid and recharge the capacitors.

In practice, we recommend setting $A_\textrm{m}^* \approx 0.8$. This value achieves a favorable balance between output waveform quality and operational headroom. At $A_\textrm{m}^* = 0.8$, the converter employs a large fraction of the available module voltage as productive output levels and yields low harmonic distortion. At the same time, retaining approximately 20\% of unused modulation range preserves the converter's ability to transiently increase its output voltage---for example, during a step increase in reactive current demand---without immediately saturating the modulation index. A value significantly below 0.8 improves headroom but degrades output quality, while a value near unity may risk modulation saturation under transient conditions.

\subsection{Summary}
The proposed dual-loop control scheme achieves two objectives: precise regulation of reactive current exchange with the grid, and control of the modulation depth and module voltages. The inner current loop tracks a sinusoidal reference to inject the commanded reactive current, while the outer modulation-index loop adjusts the active power component to compensate for converter losses and maintain a healthy modulation depth. The control requires only two measurements: grid voltage (for phase-locked loop synchronization and current reference generation) and output current (for closed-loop current control). No module voltage sensors are needed. 

\section{Design Considerations and Application Scope}

\subsection{Operating Point Constraints}
The full elimination of module voltage sensors requires careful attention to operating limits during system design. The STATCOM output voltage follows from the grid voltage and the output current as
\begin{equation}
v_\textrm{o} = V_\textrm{g} + Z_\textrm{L} \cdot i_\textrm{o},
\end{equation}
where $Z_\textrm{L}$ is the coupling impedance (primarily inductive) and $i_\textrm{o}$ is the output current amplitude. For a converter with $N$ modules, the nominal module voltage equals $v_\textrm{o}/N$.

In combination with $|v_\textrm{o}| = A_\textrm{m} N V_{\textrm{sm}}$ and for purely inductive coupling ($Z_\textrm{L} = j\omega L_\textrm{g}$), the output current for purely reactive operation is
\begin{equation}
    i_\textrm{o} = \frac{A_\textrm{m} N V_{\textrm{sm}} - V_\textrm{g}}{\omega L_\textrm{g}},
    \label{eq:io_vsm}
\end{equation}
where $V_\textrm{g} = |\mathbf{v}_\textrm{g}|$ is the grid voltage amplitude. This relationship defines the achievable reactive current for a given module voltage and modulation depth. 

Since the module voltage must remain below the rated limit $V_{\textrm{sm,max}}$, and the grid voltage may rise to $V_{\textrm{g,max}}$, the maximum deliverable current amplitude is bounded per
\begin{equation}
    i_\textrm{o} \leq \frac{A_\textrm{m}^* N V_{\textrm{sm,max}} - V_{\textrm{g,max}}}{\omega L_\textrm{g}}.
    \label{eq:io_max}
\end{equation}
This constraint must be satisfied at the design stage by choosing $N$ and $V_{\textrm{sm,max}}$ appropriately. We can use this relationship both to select the appropriate number of modules for a given application and to establish a safe operating envelope.

\subsection{Application Scope of a Single-Phase STATCOM}
Although the fundamental operating principles of direction-selective parallelization, hardware-guaranteed voltage balancing, and sensorless control apply equally to three-phase implementations, this paper demonstrates the proposed approach using a single-phase STATCOM. We choose this setting for two reasons. First, single-phase operation presents a more demanding test case than three-phase systems: without the inherent balancing provided by symmetrical three-phase currents, all energy storage and voltage regulation burden falls on a single converter arm. Additionally, the modulation index estimation requires synchronous detection with low-pass filtering in the single-phase case, whereas three-phase systems can directly compute the amplitude from Clarke-transformed components without filtering delay. These considerations make the single-phase configuration an effective proving ground for the 4T4D topology and sensorless control strategy.
Second, single-phase STATCOM applications remain important in several domains. Railway traction systems, particularly those fed by single-phase AC at \qty{25}{\kilo\volt} or \qty{15}{\kilo\volt}, require reactive power compensation to mitigate voltage fluctuations caused by rapidly varying locomotive loads \cite{horita2010single, enderle2012d}. In various countries, island microgrids and rural distribution networks often operate with single-phase feeders where localized voltage support is essential for power quality.

\section{Simulation Validation}

We first developed a simulation model in PLECS (v4.9.8, Plexim GmbH) to validate the 4T4D-based STATCOM operation.
This simulation models a \qty{10}{\kilo\volt} single-phase system.
The shunt-connected STATCOM consists of ten modules, each with a nominal voltage range of \qtyrange{1300}{1500}{\volt}.
Each module consists of four IGBTs, four diodes, and a \qty{44}{\milli\farad} capacitor.

\subsection{Steady-State}
\begin{figure}[t]
    \centering
    \includegraphics{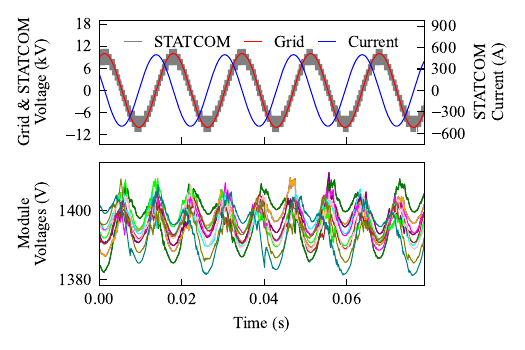}
    \caption{Simulated steady-state waveforms at \qty{500}{\ampere} current reference. The output current maintains a THD of 5.51\%. Individual module voltages are balanced around \qty{1396}{\volt} with a \qty{30}{\volt} fluctuation.}
    \label{fig:steady_state_simulation}
\end{figure}

Figure~\ref{fig:steady_state_simulation} shows the simulated steady-state waveforms with the current reference amplitude set at \qty{500}{\ampere}.
The output current achieves a total harmonic distortion (THD) of 5.51\%, and the output voltage THD is 10.66\%.
The individual module voltages remain well-regulated around \qty{1396}{\volt} with a fluctuation range of \qty{30}{\volt}.

\subsection{Load-Step Response}
\begin{figure}
    \centering
    \includegraphics{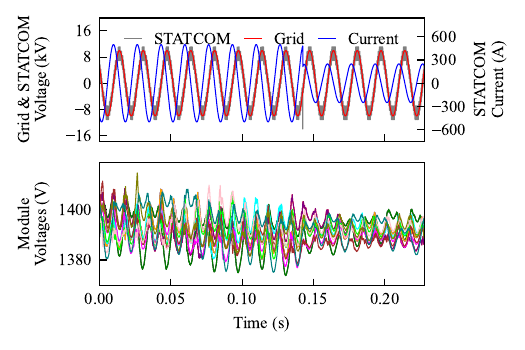}
    \caption{Simulated load step response during a current reference step from \qty{500}{\ampere} to \qty{250}{\ampere}. The output current settles within \qty{2}{\milli\second}. Individual module voltages remain balanced throughout the transient with fluctuation shrinking from \qty{39}{\volt} to \qty{30}{\volt} as the load current reduces.}
    \label{fig:step_change_simulation}
\end{figure}

Figure \ref{fig:step_change_simulation} presents the dynamics during a current step. The current reference steps from \qty{500}{\ampere} to \qty{250}{\ampere} at 140~ms.
The output current tracks the reference within \qty{1}{\milli\second} settling time.
The collective module voltage fluctuation range shrinks from \qty{39}{\volt} to \qty{30}{\volt} due to the reduction of the load current.

\subsection{Voltage Convergence}
\begin{figure}[t]
    \centering
    \includegraphics{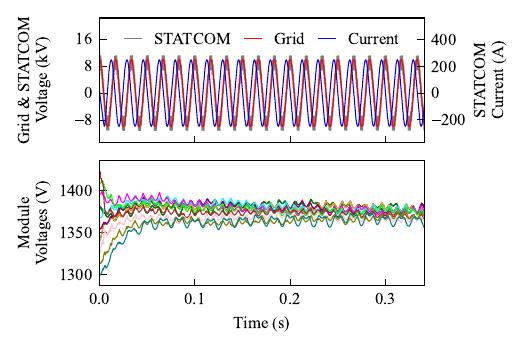}
    \caption{Simulated voltage balancing convergence while tracking a \qty{250}{\ampere} current reference. Individual module voltages started from an initial spread of $\pm 5\%$ and converged to \qty{1374}{\volt} within 150~ms.}
    \label{fig:voltage_converge_simulation}
\end{figure}
Figure \ref{fig:voltage_converge_simulation} presents the dynamics starting with an initial condition in which module voltages span a $\pm 5\%$ range, i.e., at $1360~\textrm{V} \times \{95\%, 96\%, \cdots, 104\%\}$. 
While the system closely tracks the reference current at \qty{250}{\ampere} amplitude, the voltage converges to \qty{1374}{\volt} within 150~ms without any overshoot.

\section{Experimental Platform Establishment}
\begin{figure}
    \centering
    \includegraphics[width=0.45\textwidth]{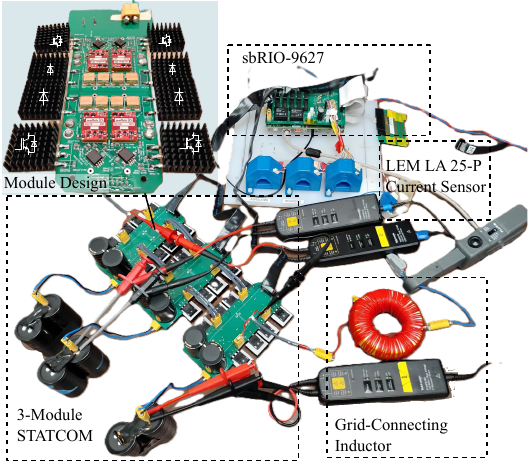}
    \caption{Experimental platform for the three-module single-phase 4T4D-based STATCOM. The converter interfaces an emulated \qty{100}{\volt}, \qty{60}{\hertz} grid through a \qty{3.58}{\milli\henry} inductor. An embedded controller implements the dual-loop control using only grid voltage and output current measurements, with no module voltage sensors.}
    \label{fig:prototype_igbt}
\end{figure}
\subsection{Platform Overview}
\begin{table}[t]
    \centering
    \caption{System Specifications}
    \label{tab:prototype_specs}
    \begin{tabular}{l l l}
    \toprule
    \textbf{Object} & \textbf{Parameter} & \textbf{Value} \\
    \midrule
    \multirow{3}{*}{Grid-Emulating Inverter} 
        & DC Voltage & \qty{100}{\volt} \\
        & Switching Frequency & \qty{5}{\kilo\hertz} \\
        & Output Frequency & \qty{60}{\hertz} \\
    \midrule
    \multirow{2}{*}{Grid Interface}
        & Connection & Single-phase \\
        & Grid Forming Inductance & \qty{3.58}{\milli\henry} \\
    \midrule
    \multirow{4}{*}{Modular STATCOM}
        & Number of Modules & 3 \\
        & Nominal Module Voltage & \qtyrange{40}{50}{\volt} \\
        & Module Capacitance & \qty{30}{\milli\farad} \\
        & Switching Frequency & \qty{5}{\kilo\hertz} \\
    \midrule
    \multirow{2}{*}{Control}
        & Controller & sbRIO-9627 \\
        & Target Modulation Index & 0.8 \\
    \bottomrule
    \end{tabular}
\end{table}

Figure~\ref{fig:prototype_igbt} illustrates the platform layout; Table~\ref{tab:prototype_specs} lists the system specifications. 
The power grid is emulated with a full-bridge inverter that operates with a dc voltage of \qty{100}{\volt} and switching rate of \qty{5}{\kilo\hertz}.
The three-module STATCOM interfaces the emulated single-phase grid through a \qty{3.58}{\milli\henry} series inductor. The system injects and absorbs reactive current in response to the reference command. The dual-loop control requires only measurements of the grid voltage and the output current; not a single module voltage sensor is used.

\subsection{Module Design and Performance}
\begin{figure}
    \centering
    \includegraphics{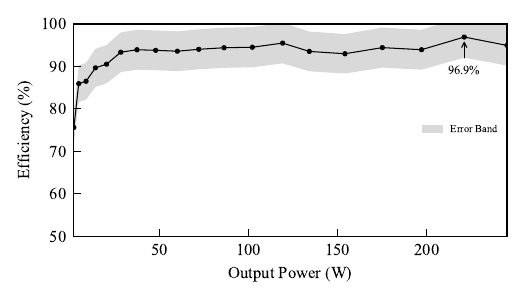}
    \caption{Efficiency curve on the module level with a \qty{100}{\volt} DC link, \qty{20}{\Omega} resistive load, and \qty{5}{\kilo\hertz} carrier frequency. The peak efficiency reaches 96.9\%. The average stays above 93\% for output powers above \qty{50}{\watt}.}
    \label{fig:single_module_efficiency}
\end{figure}
\begin{figure}
    \centering
    \includegraphics[width=0.25\textwidth]{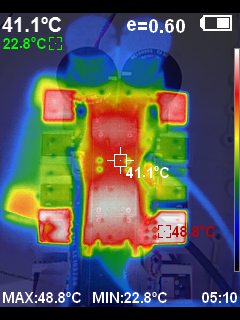}
    \caption{Thermal image of a single 4T4D module at steady state with \qty{250}{\watt} output. All critical components remain below \qty{50}{\celsius}. The highest temperature of \qty{48.8}{\celsius} occurs at an IGBT.}
    \label{fig:thermal_photo}
\end{figure}
Each 4T4D module contains four IGBTs, four silicon diodes, and a \qty{30}{\milli\farad} electrolytic capacitor. 
Modules are connected with AWG-12 wires without additional discrete inductors or magnetics. 

We tested individual module power performance by operating them as inverters, i.e., we disabled the parallel state and operated A\,\&\,B as well as C\,\&\,D \textit{in unison}.
We measured the efficiency curve for the DC-link voltage set to \qty{100}{\volt} and a \qty{20}{\Omega} resistive load (Fig.~\ref{fig:single_module_efficiency}).
The output power is derived from the voltage measured by the oscilloscope probe (THDP0100, Tektronix), and the input power is read from the power supply (HP 6030A) display. The results carry an approximate 5\% relative uncertainty due to measurement precision.
The inverter mode is controlled with a sinusoidal PWM scheme with a carrier frequency of \qty{5}{\kilo\hertz} and a dead time of \qty{500}{\nano\second}.
The efficiency peaks at 96.9\% and remains above 93\% for output power above \qty{50}{\watt}.

We monitored the thermal performance at \qty{250}{\watt} steady state with a thermal imaging camera (HT-19, Hti). All critical components remain below \qty{50}{\celsius} with active air cooling (Fig. \ref{fig:thermal_photo}). The temperature rise of the IGBTs is more significant than that of the diodes.
The highest temperature of \qty{48.8}{\celsius} occurs at the IGBT of Bridge A; all diodes stay below \qty{35}{\celsius}.

\subsection{Control Platform}
An sbRIO-9627 embedded controller with field-programmable gate array and signal processing cores (National Instruments) controls the STATCOM prototype. The control scheme follows Figure \ref{fig:control_diagram}.
Gate signals are generated by the FPGA and distributed to individual module gate drivers. 
The grid current is measured through a Hall sensor (LEM, LA 25-P) and then filtered in the embedded controller.

The default target modulation index is set to 0.8.
The low-pass filter parameter $\alpha$ for modulation index extraction as in \eqref{equ:lpf_iq} is set to \num{5e-5}, which attains an approximate bandwidth of \qty{8.3}{\hertz}, for a fast response without inducing voltage overshoots.

\section{Experimental Results}

\subsection{Steady State}
\begin{figure}
    \centering
    \includegraphics{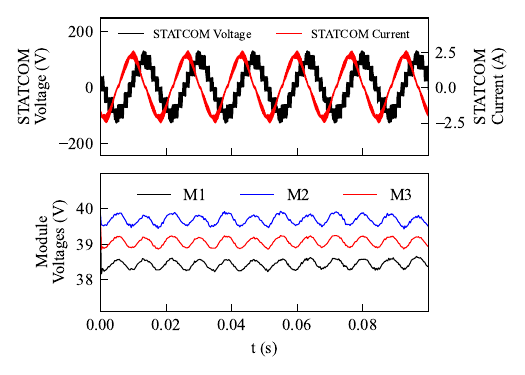}
    \caption{Experimental steady-state waveforms with seven-level output voltage synchronized to the \qty{60}{\hertz} grid. The system regulates the output current to \qty{2.5}{\ampere} with a \qty{90}{\degree} phase lag. The three module voltages stay balanced at \qty{38.4}{\volt}, \qty{39.0}{\volt}, and \qty{39.7}{\volt}.}
    \label{fig:experimental_results_steady_state}
\end{figure}

\begin{comment}
\begin{figure}[t]
    \centering
    \includegraphics{img/experimental_results_steady_state_spectrum.pdf}
    \caption{Harmonic spectrum of the output current at rated operation (\qty{2.5}{\ampere}, \qty{60}{\hertz}), yielding a THD of 7.1\%.}
    \label{fig:experimental_results_steady_state_spectrum}
\end{figure}
\end{comment}

Figure~\ref{fig:experimental_results_steady_state} graphs the steady-state waveforms of the STATCOM.
The output voltage features a seven-level staircase waveform synchronized to the \qty{60}{\hertz} grid.
The output current is regulated to a \qty{2.5}{\ampere} amplitude with a \qty{90}{\degree} phase lag relative to the voltage.

The modules maintain their voltages at respectively \qty{38.4}{\volt}, \qty{39.0}{\volt}, and \qty{39.7}{\volt}.
The voltage deviation between modules stems from the diode forward voltage in the parallelization paths.

\subsection{Load-Step Test}
\begin{figure}
    \centering
    \includegraphics{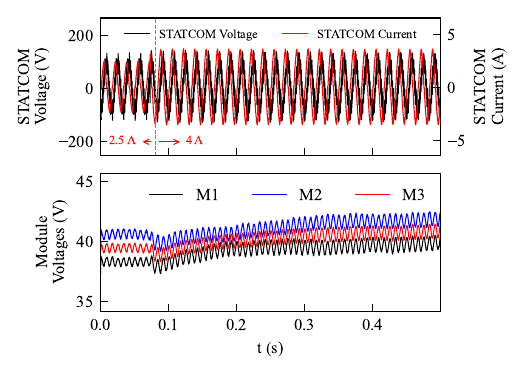}
    \caption{System response to a reactive current step from \qty{2.5}{\ampere} to \qty{4.0}{\ampere}. The current tracks the new reference within milliseconds. Module voltages experience an instantaneous drop due to the increased current demand, then re-stabilize within 200\,{}ms at a \qty{1.4}{\volt} higher level to maintain the target modulation depth of 0.8.}
    \label{fig:experimental_results_step_change_of_current}
\end{figure}

\begin{figure}
    \centering
    \includegraphics{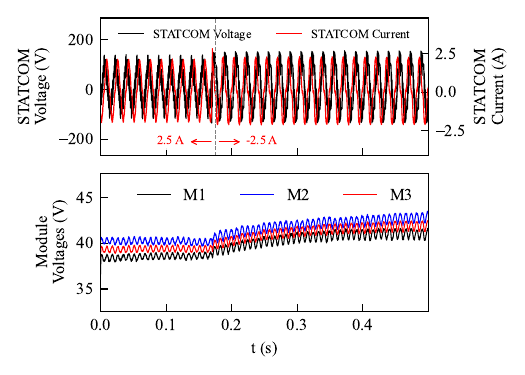}
    \caption{System response to a reactive current direction reversal at 175\,{}ms, switching from lagging to leading. The current tracks the new reference within a quarter of the output cycle. Module voltages gradually increase to a new steady state \qty{2.5}{\volt} higher on average.}
    \label{fig:experimental_results_reverse_change_of_current}
\end{figure}

Figure~\ref{fig:experimental_results_step_change_of_current} represents the system dynamics during a reactive current step from \qty{2.5}{\ampere} to \qty{4.0}{\ampere}.
Upon the command change at 8~ms, the current tracks the new reference within a few milliseconds.
Meanwhile, module voltages instantly drop because of the increased output current and stabilize at a new steady state within the next 200\,{}ms.
The new steady-state module voltage increased by \qty{1.4}{\volt} to maintain a modulation index of 0.8.

Figure~\ref{fig:experimental_results_reverse_change_of_current} demonstrates the system dynamics during reactive current direction reversal.
At 175~ms, the STATCOM current reverses direction, i.e., from lagging to leading the grid voltage.
The current settled to the new reference within a quarter of the output cycle.
Meanwhile, module voltages gradually increased to the new steady state.
The new steady-state module voltage increased by \qty{2.5}{\volt} on average.

\subsection{Voltage Regulation Validation}
\begin{figure}
    \centering
    \includegraphics{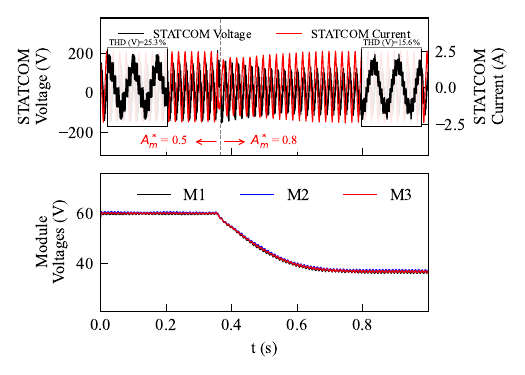}
    \caption{System response to a step change in target modulation depth $A_\textrm{m}^*$ from 0.5 to 0.8 at 365\,{}ms. At $A_\textrm{m}^* = 0.5$, module voltages stabilize at \qty{60.1}{\volt} with a five-level output waveform and current THD of 10.16\%. After the step, module voltages settle at \qty{38}{\volt} within 200\,{}ms, the output exhibits more levels, and current THD improves to 7.34\%, demonstrating that a higher $A_\textrm{m}^*$ drives module voltages lower and yields better output quality.}
    \label{fig:experimental_results_step_change_of_amref}
\end{figure}
Figure \ref{fig:experimental_results_step_change_of_amref} demonstrates the modulation-index regulation principle.
The trial starts with an $A_\textrm{m}^*$ value of 0.5; the converter produces only five output levels, with module voltages stabilized around \qty{60.1}{\volt} to deliver the \qty{2.5}{\ampere} current reference.

At time 365\,{}ms, $A_\textrm{m}^*$ is stepped to 0.8.
According to \eqref{eq:vsm_estimation}, a higher $A_\textrm{m}^*$ reduces the module voltage.
Within 200\,{}ms, the system settles at a module voltage around \qty{38}{\volt}, and the output voltage exhibits additional levels due to the increased modulation depth.
The voltage THD is 25.28\% at $A_\textrm{m}^* = 0.5$ and 15.57\% at $A_\textrm{m}^* = 0.8$.
The current THD is 10.16\% at $A_\textrm{m}^* = 0.5$ and 7.34\% at $A_\textrm{m}^* = 0.8$.
This confirms that a higher $A_\textrm{m}^*$ yields lower module voltage and better output quality.

\section{Conclusion}
Previous modular STATCOM designs faced a fundamental trade-off: conventional topologies avoid sensor problems by relying on software-based balancing that demands one voltage sensor per module, whereas parallel-capable topologies achieve hardware-guaranteed balancing but require twice the transistors. The latest parallel-capable STATCOM research still retains one sensor per arm. This paper breaks that trade-off by combining the 4T4D topology with a fully sensorless dual-loop control strategy. The 4T4D module achieves bidirectional parallelization and bipolar output with only four transistors per module---equal to a conventional full-bridge---and provides hardware-guaranteed voltage balancing through direct charge equilibration. The proposed outer modulation-index loop eliminates all module voltage sensors by inferring the module voltage from the amplitude of the current controller output and regulates active power exchange to maintain a target modulation depth. The inner current loop provides fast and precise reactive current control. The complete system requires only grid voltage and output current measurements.

Simulations and experimental results on a prototype confirm robust operation across steady-state, transient, and grid disturbance conditions. This work demonstrates the first modular STATCOM that simultaneously achieves minimum transistor count and complete elimination of module voltage sensors. 

\bibliographystyle{Bibliography/IEEEtranTIE}
\bibliography{Bibliography/IEEEabrv,Bibliography/main}\ %IEEEabrv instead of IEEEfull

@ARTICLE{goetz2015modular,
  author={Goetz, Stefan M. and Peterchev, Angel V. and Weyh, Thomas},
  journal={IEEE Transactions on Power Electronics},
  title={Modular Multilevel Converter With Series and Parallel Module Connectivity: Topology and Control},
  year={2015},
  volume={30},
  number={1},
  pages={203-215},
  doi={10.1109/TPEL.2014.2310225}}

@ARTICLE{fang2021review,
  author={Fang, Jingyang and Blaabjerg, Frede and Liu, Steven and Goetz, Stefan M.},
  journal={IEEE Transactions on Power Electronics},
  title={A Review of Multilevel Converters With Parallel Connectivity},
  year={2021},
  volume={36},
  number={11},
  pages={12468-12489},
  doi={10.1109/TPEL.2021.3075211}}

@INPROCEEDINGS{goetz2016sensorless,
  author={Goetz, Stefan M. and Li, Zhongxi and Peterchev, Angel V. and Liang, Xinyu and Zhang, Chengduo and Lukic, Srdjan M.},
  booktitle={2016 IEEE Applied Power Electronics Conference and Exposition (APEC)},
  title={Sensorless scheduling of the modular multilevel series-parallel converter: enabling a flexible, efficient, modular battery},
  year={2016},
  volume={},
  number={},
  pages={2349-2354},
  doi={10.1109/APEC.2016.7468193}}

@ARTICLE{fang2021multilevel,
  author={Fang, Jingyang and Li, Zhongxi and Goetz, Stefan M.},
  journal={IEEE Transactions on Power Electronics},
  title={Multilevel Converters With Symmetrical Half-Bridge Submodules and Sensorless Voltage Balance},
  year={2021},
  volume={36},
  number={1},
  pages={447-458},
  doi={10.1109/TPEL.2020.3000469}}

@ARTICLE{li2019module,
  author={Li, Zhongxi and Lizana F., Ricardo and Sha, Sha and Yu, Zhujun and Peterchev, Angel V. and Goetz, Stefan M.},
  journal={IEEE Transactions on Power Electronics},
  title={Module Implementation and Modulation Strategy for Sensorless Balancing in Modular Multilevel Converters},
  year={2019},
  volume={34},
  number={9},
  pages={8405-8416},
  doi={10.1109/TPEL.2018.2886147}}

@ARTICLE{tashakor2023voltage,
  author={Tashakor, Nima and Keshavarzi, Davood and Iraji, Farzad and Banana, Shady and Goetz, Stefan},
  journal={IEEE Transactions on Power Electronics},
  title={Voltage Estimation in Combination With Level-Adjusted Phase-Shifted-Carrier Modulation ({LA-PSC}) for Sensorless Balancing of Diode-Clamped Modular Multilevel Converters ({MMC}s)},
  year={2023},
  volume={38},
  number={4},
  pages={4267-4278},
  doi={10.1109/TPEL.2022.3226421}}

@ARTICLE{fang2021reduction,
  author={Fang, Jingyang and Yang, Shunfeng and Wang, Haiyu and Tashakor, Nima and Goetz, Stefan M.},
  journal={IEEE Transactions on Power Electronics},
  title={Reduction of {MMC} Capacitances Through Parallelization of Symmetrical Half-Bridge Submodules},
  year={2021},
  volume={36},
  number={8},
  pages={8907-8918},
  doi={10.1109/TPEL.2021.3049389}}

@ARTICLE{goetz2017control,
  author={Goetz, Stefan M. and Li, Zhongxi and Liang, Xinyu and Zhang, Chengduo and Lukic, Srdjan M. and Peterchev, Angel V.},
  journal={IEEE Transactions on Power Electronics},
  title={Control of Modular Multilevel Converter With Parallel Connectivity—Application to Battery Systems},
  year={2017},
  volume={32},
  number={11},
  pages={8381-8392},
  doi={10.1109/TPEL.2016.2645884}}

@ARTICLE{li2021reduced,
  author={Li, Zhongxi and Motwani, Jayesh K. and Zeng, Zhiyong and Lukic, Srdjan M. and Peterchev, Angel V. and Goetz, Stefan M.},
  journal={IEEE Transactions on Industrial Electronics},
  title={A Reduced Series/Parallel Module for Cascade Multilevel Static Compensators Supporting Sensorless Balancing},
  year={2021},
  volume={68},
  number={1},
  pages={15-24},
  doi={10.1109/TIE.2020.2965470}}

@article{tashakor2020modular,
  title={Modular multilevel converter with sensorless diode-clamped balancing through level-adjusted phase-shifted modulation},
  author={Tashakor, Nima and Kilictas, Muhammet and Bagheri, Ehsan and Goetz, Stefan},
  journal={IEEE Transactions on Power Electronics},
  volume={36},
  number={7},
  pages={7725--7735},
  year={2020},
  publisher={IEEE}
}

@article{eroglu2022critical,
  author={Eroğlu, Fatih and Vural, Ahmet Mete},
  journal={Global Power, Energy and Communication Conference (GPECOM)},
  title={A Critical Review on State-of-Charge Balancing Methods in Multilevel Converter Based Battery Storage Systems},
  year={2022},
  volume={4},
  pages={14-19},
  doi={10.1109/GPECOM55404.2022.9815810}}

@ARTICLE{wu2021comprehensive,
  author={Wu, Mingzhe and Li, Yun Wei and Konstantinou, Georgios},
  journal={IEEE Transactions on Power Electronics},
  title={A Comprehensive Review of Capacitor Voltage Balancing Strategies for Multilevel Converters Under Selective Harmonic Elimination PWM},
  year={2021},
  volume={36},
  number={3},
  pages={2748-2767},
  doi={10.1109/TPEL.2020.3012915}}

@ARTICLE{milovanovic2021comprehensive,
  author={Milovanović, Stefan and Dujić, Dražen},
  journal={IEEE Transactions on Power Electronics},
  title={Comprehensive Comparison of Modular Multilevel Converter Internal Energy Balancing Methods},
  year={2021},
  volume={36},
  number={8},
  pages={8962-8977},
  doi={10.1109/TPEL.2021.3052607}}

@article{konstantinou2018estimation,
  author={Konstantinou, Georgios and Wickramasinghe, Harith R. and Townsend, Christopher D. and Ceballos, Salvador and Pou, Josep},
  journal={International Conference on Power and Energy Systems (ICPES)},
  title={Estimation Methods and Sensor Reduction in Modular Multilevel Converters: A Review},
  year={2018},
  volume={8},
  pages={23-28},
  doi={10.1109/ICPESYS.2018.8626987}}

@article{ma2018enhanced,
  author={Ma, Zhan and Hao, Tianqu and Gao, Feng and Li, Nan and Gu, Xin},
  journal={IEEE Applied Power Electronics Conference and Exposition (APEC)},
  title={Enhanced {SOH} balancing method of {MMC} battery energy storage system with cell equalization capability},
  year={2018},
  pages={3591-3597},
  doi={10.1109/APEC.2018.8341622}}

@ARTICLE{dekka2017evolution,
  author={Dekka, Apparao and Wu, Bin and Fuentes, Ricardo Lizana and Perez, Marcelo and Zargari, Navid R.},
  journal={IEEE Journal of Emerging and Selected Topics in Power Electronics},
  title={Evolution of Topologies, Modeling, Control Schemes, and Applications of Modular Multilevel Converters},
  year={2017},
  volume={5},
  number={4},
  pages={1631-1656},
  doi={10.1109/JESTPE.2017.2742938}}

@inproceedings{zhang2024frequency,
  title={Frequency-Dependent Impedance Variation in Multilevel Converters with Parallel Connectivity},
  author={Zhang, Jinshui and Peterchev, Angel V and Goetz, Stefan M},
  booktitle={2024 IEEE Applied Power Electronics Conference and Exposition (APEC)},
  pages={2337--2341},
  year={2024},
  organization={IEEE}
}

@inproceedings{yin2018modular,
  title={Modular multilevel converter with capacitor voltage self-balancing using reduced number of voltage sensors},
  author={Yin, Taiyuan and Wang, Yue and Wang, Xiaolei and Yin, Shiyuan and Sun, Shumin and Li, Guanglei},
  booktitle={2018 International Power Electronics Conference (IPEC-Niigata 2018-ECCE Asia)},
  pages={1455--1459},
  year={2018},
  organization={IEEE}
}

@article{liu2016novel,
  title={A novel STATCOM based on diode-clamped modular multilevel converters},
  author={Liu, Xiangdong and Lv, Jingliang and Gao, Congzhe and Chen, Zhen and Chen, Si},
  journal={IEEE Transactions on Power Electronics},
  volume={32},
  number={8},
  pages={5964--5977},
  year={2016},
  publisher={IEEE}
}

@inproceedings{zheng2017medium,
  title={A medium-voltage motor drive based on diode-clamped modular multilevel converters},
  author={Zheng, Tong and Gao, Congzhe and Liao, Xiaozhong and Liu, Xiangdong and Sun, Baiyan and Lv, Jingliang},
  booktitle={2017 20th International Conference on Electrical Machines and Systems (ICEMS)},
  pages={1--6},
  year={2017},
  organization={IEEE}
}

@ARTICLE{tashakor2021switch,
  author={Tashakor, Nima and Kılıçtaş, Muhammet and Fang, Jingyang and Goetz, Stefan M.},
  journal={IEEE Transactions on Industrial Electronics},
  title={Switch-Clamped Modular Multilevel Converters With Sensorless Voltage Balancing Control},
  year={2021},
  volume={68},
  number={10},
  pages={9586-9597},
  doi={10.1109/TIE.2020.3028814}}

@ARTICLE{tang2016compact,
  author={Tang, Yuan and Chen, Minjie and Ran, Li},
  journal={IEEE Transactions on Power Electronics},
  title={A Compact {MMC} Submodule Structure With Reduced Capacitor Size Using the Stacked Switched Capacitor Architecture},
  year={2016},
  volume={31},
  number={10},
  pages={6920-6936},
  doi={10.1109/TPEL.2015.2511189}}

@article{merlin2015cell,
  author = {Merlin, Michaël M. C. and Green, Tim C.},
  title = {Cell capacitor sizing in multilevel converters: cases of the modular multilevel converter and alternate arm converter},
  journal = {IET Power Electronics},
  volume = {8},
  number = {3},
  pages = {350-360},
  doi = {https://doi.org/10.1049/iet-pel.2014.0328},
  year = {2015}
}

@inproceedings{zhang2024direction,
  title={Direction-Selective Parallel Module Structure for Cascaded Bridge and Modular Multilevel Converters with Minimum Transistor Count},
  author={Zhang, Jinshui and Goetz, Stefan M},
  booktitle={IECON 2024-50th Annual Conference of the IEEE Industrial Electronics Society},
  pages={1--6},
  year={2024},
  organization={IEEE}
}

@article{hingorani2002flexible,
  title={Flexible AC transmission},
  author={Hingorani, Narain G},
  journal={IEEE spectrum},
  volume={30},
  number={4},
  pages={40--45},
  year={2002},
  publisher={IEEE}
}

@article{peng2017flexible,
  title={Flexible AC transmission systems (FACTS) and resilient AC distribution systems (RACDS) in smart grid},
  author={Peng, Fang Z},
  journal={Proceedings of the IEEE},
  volume={105},
  number={11},
  pages={2099--2115},
  year={2017},
  publisher={IEEE}
}

@article{singh2009static,
  title={Static synchronous compensators (STATCOM): a review},
  author={Singh, Bhim and Saha, R and Chandra, Ambrish and Al-Haddad, Kamal},
  journal={IET power electronics},
  volume={2},
  number={4},
  pages={297--324},
  year={2009},
  publisher={IET}
}

@article{sharma2023comprehensive,
  title={A comprehensive review on STATCOM: paradigm of modeling, control, stability, optimal location, integration, application, and installation},
  author={Sharma, Sandeep and Gupta, Saket and Zuhaib, Mohd and Bhuria, Vijay and Malik, Hasmat and Almutairi, Abdulaziz and Afthanorhan, Asyraf and Hossaini, Mohammad Asef},
  journal={IEEE Access},
  volume={12},
  pages={2701--2729},
  year={2023},
  publisher={IEEE}
}

@inproceedings{horita2010single,
  title={Single-phase STATCOM for feeding system of Tokaido Shinkansen},
  author={Horita, Yasuhisa and Morishima, Naoki and Kai, Masahiko and Onishi, Mitsuru and Masui, Takeshi and Noguchi, Masaki},
  booktitle={The 2010 International Power Electronics Conference-ECCE ASIA-},
  pages={2165--2170},
  year={2010},
  organization={IEEE}
}

@inproceedings{enderle2012d,
  title={D-STATCOM applied to single-phase distribution networks: Modeling and control},
  author={Enderle, Taciana Paula and da Silva, Guilherme S and Fischer, Cl{\'e}cio and Beltrame, Rafael Concatto and Schuch, Luciano and Montagner, Vinicius Foletto and Rech, Cassiano},
  booktitle={IECON 2012-38th Annual Conference on IEEE Industrial Electronics Society},
  pages={321--326},
  year={2012},
  organization={IEEE}
}

@article{AMIRREZAI2023109253,
title = {Feasibility study of incorporating static compensators in distribution networks containing distributed generation considering system power factor},
journal = {Electric Power Systems Research},
volume = {219},
pages = {109253},
year = {2023},
issn = {0378-7796},
doi = {https://doi.org/10.1016/j.epsr.2023.109253},
url = {https://www.sciencedirect.com/science/article/pii/S0378779623001426},
author = {Masoud Amirrezai and Hamid Rezaie and Stefan M. Goetz}
}

@ARTICLE{10721336,
  author={Zhang, Jinshui and Wang, Boshuo and Tian, Xiaoyang and Peterchev, Angel V. and Goetz, Stefan M.},
  journal={IEEE Transactions on Industrial Electronics}, 
  title={Analytical Model and Planar Magnetic Solution for Parallelization Surges in Switched-Capacitor and Series/Parallel Multilevel Circuits}, 
  year={2025},
  volume={72},
  number={5},
  pages={4742-4750},
  doi={10.1109/TIE.2024.3472297}}

\end{document}